\documentclass[screen,authorversion]{acmart} %

\usepackage{tabularx} 
\usepackage{graphicx}
\usepackage{subcaption}
\usepackage{booktabs}
\usepackage{caption}
\usepackage{siunitx}

\newcommand*{\SuperScriptSameStyle}[1]{%
  \ensuremath{%
    \mathpalette\SuperScriptSameStyleAux{#1}%
  }%
}
\newcommand*{\SuperScriptSameStyleAux}[2]{%
  {}^{#1#2}%
}

\newcommand*{\threeS}{\SuperScriptSameStyle{*{*}*}}

\AtBeginDocument{%
  }

\copyrightyear{2025}
\acmYear{2025}
\setcopyright{none}
\acmConference[LAK 2025]{LAK25: The 15th International Learning Analytics and Knowledge Conference}{March 03--07, 2025}{Dublin, Ireland}
\acmBooktitle{LAK25: The 15th International Learning Analytics and Knowledge Conference (LAK 2025), March 03--07, 2025, Dublin, Ireland}
\acmDOI{10.1145/3706468.3706505}
\acmISBN{979-8-4007-0701-8/25/03}

\begin{document}

\title[Does the Doer Effect Exist Beyond WEIRD Populations?]{Does the Doer Effect Exist Beyond WEIRD Populations? Toward Analytics in Radio and Phone-Based Learning}

\author{Darren Butler}
\email{ddbutler@cs.cmu.edu}
\orcid{https://orcid.org/0000-0002-2712-2863}
\affiliation{%
  \institution{Carnegie Mellon University}
  \city{Pittsburgh}
  \state{PA}
  \country{USA}
}

\author{Conrad Borchers}
\email{cborcher@cs.cmu.edu}
\orcid{https://orcid.org/0000-0003-3437-8979}
\affiliation{%
  \institution{Carnegie Mellon University}
  \city{Pittsburgh}
  \state{PA}
  \country{USA}
}

\author{Michael W. Asher}
\email{masher@andrew.cmu.edu}
\orcid{https://orcid.org/0000-0002-1006-8813}
\affiliation{%
  \institution{Carnegie Mellon University}
  \city{Pittsburgh}
  \state{PA}
  \country{USA}
}

\author{Yongmin Lee}
\email{yongminl@andrew.cmu.edu}
\orcid{https://orcid.org/0009-0006-7632-7511}
\affiliation{%
  \institution{Carnegie Mellon University}
  \city{Pittsburgh}
  \state{PA}
  \country{USA}
}

\author{Sonya Karnataki}
\email{skarnata@andrew.cmu.edu}
\orcid{https://orcid.org/0009-0000-7149-6689}
\affiliation{%
  \institution{Carnegie Mellon University}
  \city{Pittsburgh}
  \state{PA}
  \country{USA}
}

\author{Sameeksha Dangi}
\email{sameeksd@andrew.cmu.edu}
\orcid{https://orcid.org/0009-0004-4618-0251}
\affiliation{%
  \institution{Carnegie Mellon University}
  \city{Pittsburgh}
  \state{PA}
  \country{USA}
}

\author{Samyukta Athreya}
\email{ssathrey@andrew.cmu.edu}
\orcid{https://orcid.org/0009-0004-6297-1539}
\affiliation{%
  \institution{Carnegie Mellon University}
  \city{Pittsburgh}
  \state{PA}
  \country{USA}
}

\author{John Stamper}
\email{jstamper@andrew.cmu.edu}
\orcid{https://orcid.org/0000-0002-2291-1468}
\affiliation{%
  \institution{Carnegie Mellon University}
  \city{Pittsburgh}
  \state{PA}
  \country{USA}
}

\author{Amy Ogan}
\email{aeo@andrew.cmu.edu}
\orcid{https://orcid.org/0000-0003-2671-6149}
\affiliation{%
  \institution{Carnegie Mellon University}
  \city{Pittsburgh}
  \state{PA}
  \country{USA}
}

\author{Paulo F. Carvalho}
\email{pcarvalh@andrew.cmu.edu}
\orcid{https://orcid.org/0000-0002-0449-3733}
\affiliation{%
  \institution{Carnegie Mellon University}
  \city{Pittsburgh}
  \state{PA}
  \country{USA}
}

\renewcommand{\shortauthors}{Butler, et al.}

\begin{abstract}
The Doer Effect states that completing more active learning activities, like practice questions, is more strongly related to positive learning outcomes than passive learning activities, like reading, watching, or listening to course materials. 
Although broad, most evidence has emerged from practice with tutoring systems in Western, Industrialized, Rich, Educated, and Democratic (WEIRD) populations in North America and Europe. 
Does the Doer Effect generalize beyond WEIRD populations, where learners may practice in remote locales through different technologies? 
Through learning analytics, we provide evidence from N = 234 Ugandan students answering multiple-choice questions via phones and listening to lectures via community radio. 
Our findings support the hypothesis that active learning is more associated with learning outcomes than passive learning. We find this relationship is weaker for learners with higher prior educational attainment. Our findings motivate further study of the Doer Effect in diverse populations. We offer considerations for future research in designing and evaluating contextually relevant active and passive learning opportunities including leveraging familiar technology, increasing the number of practice opportunities, and aligning multiple data sources.
\end{abstract}



\maketitle

\section{Introduction}
The urban population of the Majority World, that is, the countries representing the majority of the world's inhabitants, is projected to reach 3.75 billion by 2025 \cite{smitUrbanizationGlobalSouth2021}. Learning Analytics aims to accomplish the mission of `understanding and optimizing learning and the environments in which it occurs' \cite{siemensLearningAnalyticsEducational2012}. However, since 58\% of Learning Analytics studies focus on the Minority World \cite{baekLearningAnalyticsComparison2024}, it is important to evaluate the generalizability of our findings across populations and tools.

The \textit{Doer Effect} is the finding that completing more active learning activities, like practice questions, is more strongly related to positive learning outcomes than passive learning activities, like reading, watching, or listening to course materials \cite{vancampenhoutDoerEffectScale2023, koedingerDoerEffectCausal2016}. The COPES model of self-regulated learning suggests that learners construct knowledge influenced by operations and tools \cite{gasevicLetsNotForget2015}. How the construction of knowledge in the Doer Effect differs between learner demographics and technologies is still an open question. Previous Doer Effect findings largely come from western, educated, industrialized, rich, and democratic (WEIRD) populations \cite{henrichMostPeopleAre2010}. For example, Van Campenhout et al. \cite{vancampenhoutDoerEffectScale2023} found the Doer Effect in data from the Open Learning Initiative, which largely serves Western college students through tutored practice in graphical, point-and-click interfaces. \cite{vancampenhoutDoerEffectScale2023}. In contrast, Moloo et al. developed an “audio MOOC,” which served South Asian semiliterate fishermen with audio on low-end mobile phones.

Given the large potential variability within the global learner population and across educational technologies, it is important to evaluate whether a Doer Effect exists beyond WEIRD \cite{henrichMostPeopleAre2010} learning environments. 
We aim to fill this gap by testing for the presence of the Doer Effect in a sample of non-WEIRD learners, using basic technology (i.e. radio and mobile phones) in a mobile learning course for STEM education. We test common hypotheses about the Doer Effect, including independence across prior knowledge levels and treatment interactions with prior instruction \cite{carvalhoSkipReadingAssignment2024}. We aim to answer the following research questions:
\begin{enumerate}
    \item[\textbf{RQ1}] How does learner final exam performance change with the number of new practice questions answered? 
    \item[\textbf{RQ2}] How does performance change with practice, considering different education levels? 
    \item[\textbf{RQ3}] How does performance change with practice and prior listening to radio lecture?  Is there a Doer~Effect?
\end{enumerate}

We investigate \textbf{RQ1} to describe the relationship between practice and learning outcomes in a non-WEIRD sample. We further investigate \textbf{RQ2} the variability within a non-WEIRD population \cite{apicellaWEIRDReviewLast2020} to address the open question of the role of prior knowledge in practice behavior \cite{buchinRetrievalbasedLearningPrior2022}.  
Finally, \textbf{RQ3} addresses the open question of the role of passive learning such as listening to lectures in facilitating learning outcomes and how it compares to active learning practice to create a Doer Effect \cite{koedingerLearningNotSpectator2015}. 

\section{Background}
Previous studies of the doer effect compare learning by doing and learning through lecture or reading the materials \cite{vancampenhoutDoerEffectScale2023, koedingerLearningNotSpectator2015, koedingerDoerEffectCausal2016}. Learning-by-doing refers to active learning through cognitive tasks such as information retrieval when answering multiple-choice  \cite{koedingerLearningNotSpectator2015}. Learning-by-lecture refers to passive learning from declarative information by reading, listening to, or watching expository materials like videos or lecture slides \cite{koedingerLearningNotSpectator2015}. 
These studies have aimed to identify learning methods that yield better learning outcomes. 
Through correlational and causal models, prior work found, across large online courses and within units of singular courses, learning-by-doing more strongly correlated with learning outcomes than learning-by-lecture \cite{vancampenhoutDoerEffectScale2023, koedingerDoerEffectCausal2016}. 

Research has yet to explore this relationship with populations, whom may lack reliable access to smart devices capable of the operations required by Massive Open Online Courses (MOOC) platforms like Coursera or the Open Learning Initiative (OLI). 
Notable studies of the doer effect have leveraged Coursera and OLI as materials and data sources for their findings \cite{vancampenhoutDoerEffectScale2023, koedingerLearningNotSpectator2015, koedingerDoerEffectCausal2016}. 
In this paper, we extend prior findings by investigating the correlated impact of learning-by-doing and learning-by-lecture on learning outcomes in an environment where learning-by-doing occurs through basic mobile phones and learning-by-lecture occurs through community radio broadcasts. 
In contrast to the college-level material of previous works, we study a 
 mobile learning course designed for learners who may not have completed secondary education.

Simpler technologies can scale instruction. For example, Kizilcec et al. \cite{kizilcecStudentEngagementMobile2020} studied how 93,819 Kenyan students in grades 6, 9, and 12 used a mobile learning platform based on text messages, where the activity patterns suggested that students in Kenya use mobile learning to complement formal schooling, prepare for examinations, and bridge gaps in instruction. 
Mobile learning is the intersection of mobile computing and e-learning, which allows learning to occur anywhere and anytime \cite{motiwallaMobileLearningFramework2007}. Mobile learning has been used as an educational resource in difficult-to-reach populations \cite{banksMobileLearningLast2014}. Mobile learning is well positioned to leverage infrastructure in the approaching 3.75 billion people in the urbanizing world compared to online learning. In 2020, the World Bank estimated that 46\% of adult, rural Ugandans owned a phone, while only
5\% had internet access \cite{world2019uganda}. Thus, there is an opportunity and need to meet learners where they are technologically by leveraging basic telecommunication.
Similarly, radio is a scalable learning technology that has disseminated educational content to hard-to-reach and crisis-affected groups, especially at the height of the COVID-19 pandemic \cite{razaInteractiveRadioInstruction2022}. 

The WEIRD nature of Doer Effect research follows a larger trend in Learning Analytics and related fields. Baek and Doleck \cite{baekLearningAnalyticsComparison2024} found that 58\% (N = 492) of Learning Analytics studies published in 2015-2019 were drawn from WEIRD samples with 45\% originating from the United States and Spain alone. Similarly, other technology literature overrepresents WEIRD groups. Blanchard \cite{blanchardWEIRDNatureITS2012} found that across the Intelligent Tutoring Systems (ITS) and Artificial Intelligence in Education (AIED) conferences during 2002-2011, 61\% of the total studies contained samples from the United States alone. In 2016-2020, 73\% of papers in the ACM CHI Conference on Human Factors in Computing Systems used WEIRD samples \cite{linxenHowWEIRDCHI2021}.

The overrepresentation of WEIRD samples in Learning Analytics is related to the challenges of generalizability, diversity, equity, and inclusion \cite{bakerChallengesFutureEducational2019, gasevicLearningAnalyticsShould2016, mathraniPerspectivesChallengesGeneralizability2021}. Models and platforms designed with one group can fail or undermine the learning opportunities for the groups that were not represented in the development process (e.g., Baker's New York City Marfa problem \cite{bakerChallengesFutureEducational2019} and \cite{bayerLearningAnalyticsFairness2021}).  We push toward understanding generalizability and equity by assessing to what extent a population outside the typically represented group experiences the Doer Effect in everyday learning. We contribute to efforts in closing the Learning Analytics loop by demonstrating how an important principle arises in a typically hard-to-reach group of learners.

\section{Methods}
We investigate the presence of the Doer Effect in a non-WEIRD sample by estimating the impact of answering new practice questionson exam scores in a mobile learning course. The paper's authors did not participate in the course's design and implementation. A nonprofit organization facilitated the course and maintained the resulting log-trace data. The nonprofit removed all personally identifiable information from the dataset before sharing it with the authors; the IRB classified this work as an exempt study.

\subsection{Context}
The nonprofit facilitated the course between August 11, 2022, and December 28, 2022. The nonprofit advertised and facilitated learner sign-ups for the course through community radio and mobile phones in Northern Uganda. Uganda continues to face barriers to education equality comparable to similar low and middle-income countries in the Sub-Saharan African region and rural areas \cite{uchidiunoDesigningAppropriateLearning2018, oyelaran-oyeyinkaInternetDiffusionSubSaharan2005}. Barriers include long and hazardous traveling distances to the nearest school, lack of transportation, hefty school tuition fees \cite{irvinEducationalBarriersRural2012}, shortage of skilled instructors, and insufficient school infrastructure \cite{uchidiunoDesigningAppropriateLearning2018}. Additionally, subsistence farming is a primary source of income for many families in rural regions of low-income countries. Thus, child labor is often required at home, which can limit school attendance \cite{uchidiunoDesigningAppropriateLearning2018}. Given these attributes, we identify the learners in our sample as non-WEIRD on average. Participants in the course were not offered financial compensation. However, learners who completed the course received a certificate of completion.

\subsection{Course Features and Design}
\begin{figure*}[htbp]
  \centering
  \includegraphics[width=0.8\textwidth]{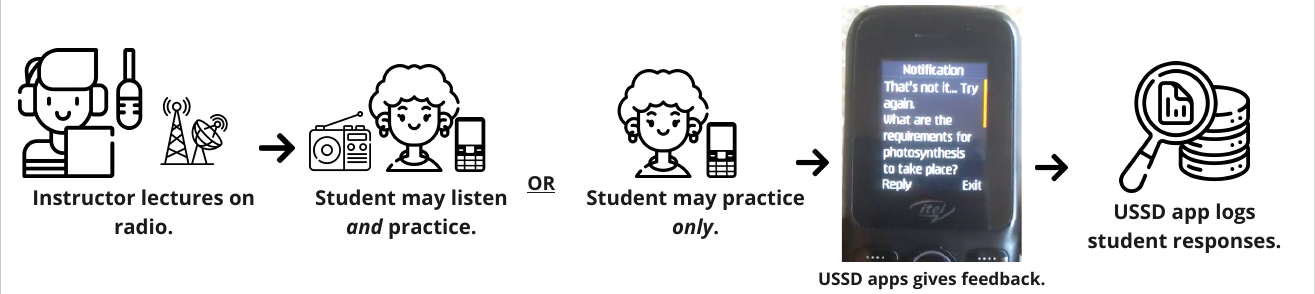}
  \caption{Instruction, Assessment, and Data Collection. A student may listen to radio instruction or practice through multiple-choice assessments on a phone. A database logs student responses via USSD messages.}
  \Description{Icons of a radio instructor, student, and a database in the cycle of 
  instruction and assessment}
  \label{fig:student-ineraction}
\end{figure*}

The nonprofit designed the course to teach out-of-school youth (those not enrolled in secondary or higher education) about the science of simple machines and engineering mindsets. The course includes active and passive learning activities. Passive learning involves listening to live radio lectures, while active learning involves phone-based multiple-choice quizzes that provide  immediate feedback for until learners select the correct answer. 
Additionally, the course features a pre-post survey and a final exam for summative evaluation. Student interactions with quizzes were logged via a USSD text message protocol. USSD is similar to SMS --- while SMS offers chat, USSD offers an interactive text menu \cite{get.africaWhatUSSD2020}. 
Learners dial codes to access the menu, navigate to practice questions, and submit answers based on options.
The resulting log data from these interactions is stored in a database. Each week featured three radio broadcasts and corresponding assessments with 2-4 questions. 
Learners chose when to engage with radio and phone content. Figure \ref{fig:student-ineraction} describes the main features of the course. 

\subsection{Dataset Processing and Measures}
We identified 4,135 total user profiles in the dataset. Of those users, 2,407 were marked as enrolled learners for this course. We calculated the following measures from the raw data.

\textbf{\textit{Doing}} is active learning represented by the total of new multiple-choice practice questions a learner answered throughout the course. 
1,753 learners attempted practice questions. Learners completed an average of 86 unique practice questions with a standard deviation of 24.8. The learners completed a median of 96 new practice questions with a minimum of 2 and a maximum of 98 over 139 days. There were 98 practice questions available. 

\textbf{\textit{Prior Listening}} is passive learning represented as the ratio of `yes' versus `no' responses learners gave when surveyed `Did you listen to the radio before answering these questions?' 
The listening question appeared at the end of 16 of the practice assessments. A total of 883 learners answered a listening question at least once. 
On average, learners responded yes versus no 35\% of the time (\textit{SD} = 37\%) with a median yes response of 0.19. 
Thus, learners often reported not listening to the radio before answering questions. Furthermore, 10\% of learners did not respond to any of the 16 listening surveys.

\textbf{\textit{Education level}} is the last level of formal education that a learner completed. In registration, learners were asked, `What is the last class you completed at school?' Learners selected categories ranging from `P7 or below' (7 years of primary education or less) to complete A levels (school-leaving qualifications used in university admissions). 
Education level serves as a proxy variable for prior knowledge. 
Most learners did not complete secondary school. Only 4.7\% of the learners completed the A Levels. The plurality of learners completed up to S4, secondary four grade level (S4: 84 (35.9\%), followed by 7 years of primary education or less (P7 or below: 60 (25.6\%)).

\textbf{\textit{Exam score}} is the number of questions a learner correctly answered on the final exam at the end of the course. 332 learners attempted the final exam, resulting in a retention rate of 13.70\% of enrolled users, which is comparable to MOOCs \cite{badaliRoleMotivationMOOCs2022}. The learners scored a mean of 10.70 points out of 15 with a standard deviation of 2.69 and a median of 11 points, range=[0, 15].

\begin{table}[htbp]
    \centering
    \caption{Summary of Learner Characteristics (\textbf{N}=234).}
    \label{tab:participant_summary}
    \begin{tabularx}{\linewidth}{lX} %
        \toprule
        \textbf{Measure} & \textbf{Overall (}N\textbf{=234)} \\ 
        \midrule
        \textbf{Doing} & Mean (\textit{SD}): 84.3 (24.80), Median [Min, Max]: 96.00 [2.00, 98.00] \\
        \addlinespace
        \textbf{Prior Listening} & Mean (SD): 0.35 (0.37), Median [Min, Max]: 0.19 [0, 1.00], Missing: 10 (4.30\%) \\
        \addlinespace
        \textbf{Education Level} & P7 or below: 60 (25.60\%), S1: 19 (8.10\%), S2: 23 (9.80\%), S3: 23 (9.80\%), S4: 4: 84 (35.90\%), S5 or above: 14 (6.00\%), Completed A level: 11 (4.70\%) \\
        \addlinespace
        \textbf{Exam Score} & Mean (\textit{SD}): 10.70 (2.69), Median [Min, Max]: 11.00 [0, 15.0] \\
        \bottomrule
    \end{tabularx}
\end{table}

\subsection{Analysis and Model}
In merging measures, our final sample contained 234 learners for whom we had all measures. Table \ref{tab:participant_summary} describes the final sample of learners. We fit linear regression models to answer each research question, interpreting the direction and strength of the associations between the explanatory variables and the outcome while accounting for important control variables \cite{jamesLinearRegression2023}. Before putting the aggregated features into regression models using standard regression model functions in R \cite{rcoreteamLanguageEnvironmentStatistical2024}, we scale all measures to a mean of 0 and standard deviation of 1 (\textit{z}-scoring) to ease interpretation of associations on a common scale of standard deviations\cite{stevensOutliersInfluentialData1984}. For RQs 2 and 3, we add interaction terms to identify how the magnitude of the Doer Effect may change when accounting for education and listening to lectures. Thus, we compose the following R formulas: 
\textbf{RQ1.} \textit{$\text{lm}( \text{exam\_score} \sim \text{doing}) \label{model-1}$};
\textbf{RQ2.} \textit{$\text{lm}( \text{exam\_score} \sim \text{doing} \times \text{education}) \label{model-2}$};
\textbf{RQ3.} \textit{$\text{lm}( \text{exam\_score} \sim \text{doing} \times \text{listeningBefore}) \label{model-3}$}.

\section{Results}

\textbf{\textit{RQ1: How does learner final exam score change with the number of new practice questions answered?}}
Doing was significantly positively related to exam scores, $CI_{95\%}$ = 0.38, $CI_{95\%}$ [0.26 – 0.50], \emph{p} <.001, indicating a positive relationship between practice engagement and exam score. 
Considering doing only, 1 $SDs$ more practice corresponded to 0.38 $SDs$ higher exam scores.
The model accounts for approximately 14.2\% of the variance in exam points between learners (Adjusted \emph{R}\textsuperscript{2} = 0.139). 

\begin{figure*}[htbp]
    \centering
    \begin{subfigure}[b]{0.40\textwidth}
        \centering
        \includegraphics[width=\linewidth, height=0.25\textheight, keepaspectratio]{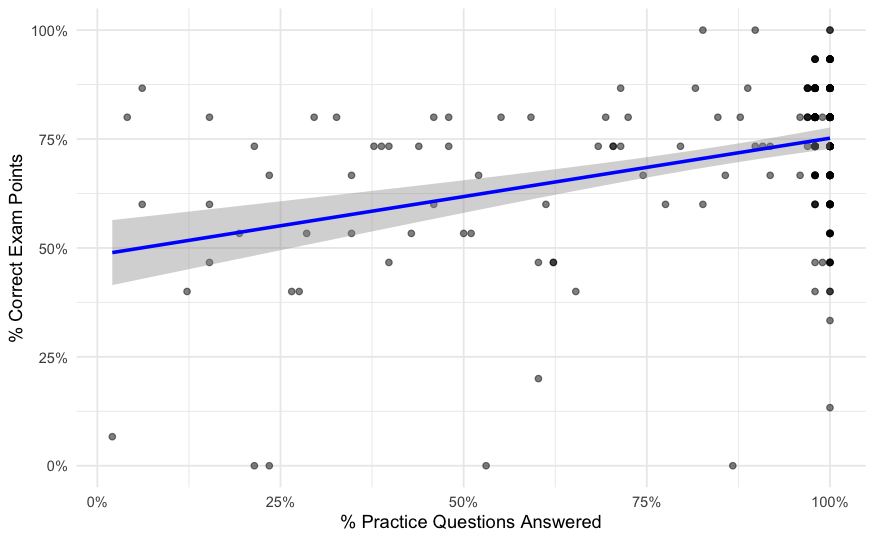}
        \caption{Between all learners}
        \label{fig:slope_main_effect}
        \Description{Plot of positive regression line fit}
    \end{subfigure}
    \hfill
    \begin{subfigure}[b]{0.50\textwidth}
        \centering
        \includegraphics[width=\linewidth, height=0.25\textheight, keepaspectratio]{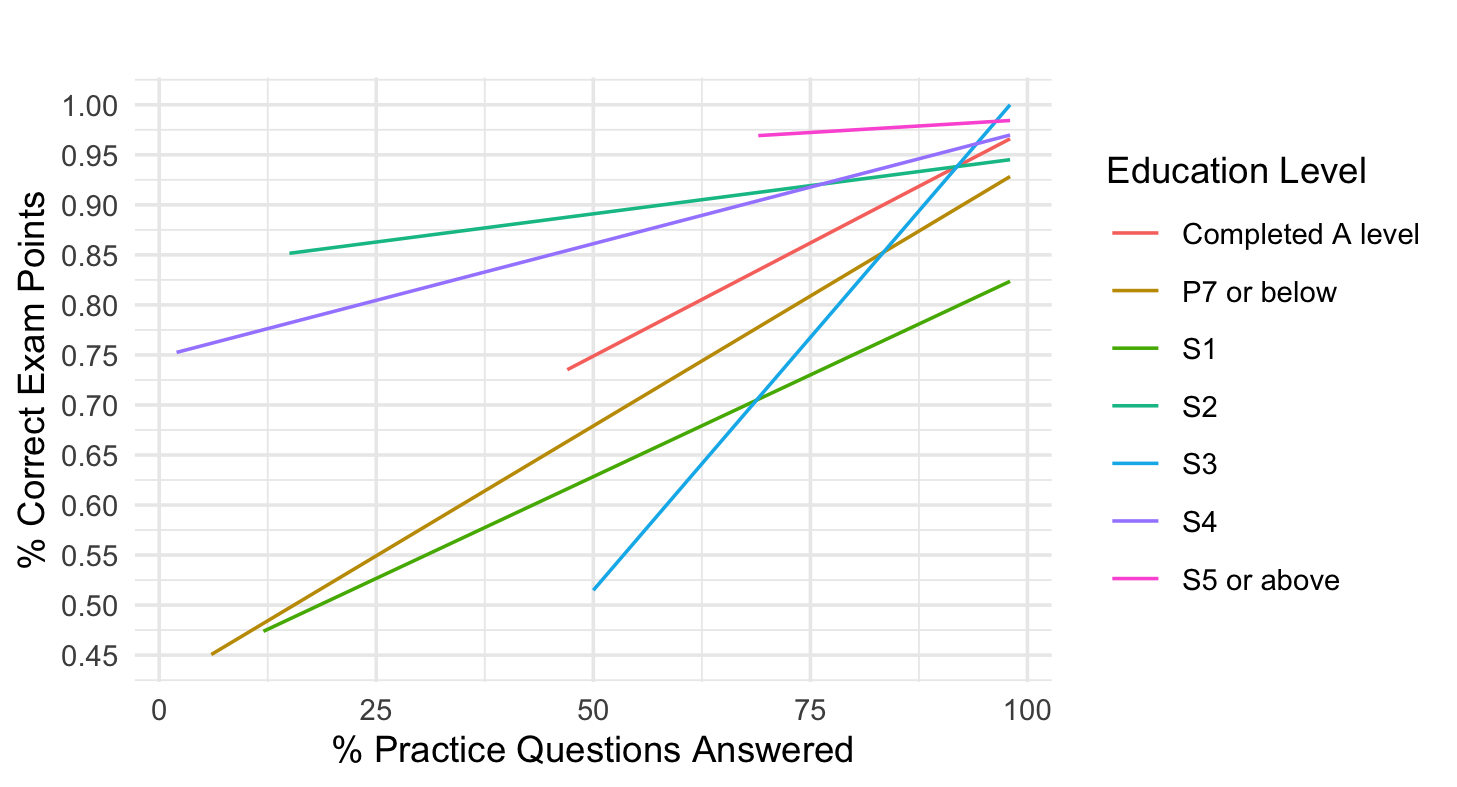}
        \caption{Between education groups}
        \label{fig:interaction}
        \Description{Interaction plot of grades}
    \end{subfigure}
    \caption{Changes in exam score across percentage of practice questions done out of the total available}
\end{figure*}

Figure \ref{fig:slope_main_effect} suggests a general trend in which exam score increases with total new practice questions answered, as indicated by the positive slope of the regression line. However, the scatter of data points in \ref{fig:slope_main_effect} shows considerable variation, suggesting that other factors may influence exam outcomes. A potential ceiling effect in practice engagement (doing) is evident, with many students (43\%) completing all practice questions. This clustering suggests that some students may have practiced more if given the opportunity. This could limit the strength of the Doer Effect measured in our sample, as variation in the doing predictor is reduced. In addition, a notable proportion of students achieved high exam grades although they did not practice much, which may be related to differences in prior knowledge.

\textbf{\textit{RQ2: How does performance change with practice, considering different education levels?}}
We test whether the Doer Effect is robust when adjusting for the level of education.
The number of unique practice questions answered was significantly related to exam scores ($CI_{95\%}$ = 0.37, $CI_{95\%}$ [0.25, 0.49], \emph{p} < 0.001), indicating a positive relationship where 1 $SDs$ more practice corresponds to 0.37 $SDs$ higher exam scores. 
No education levels show a statistically significant change in exam scores from the baseline of `Completed A Levels' ($p's$ > .05). The model accounts for approximately 19.9\% of the variance in exam points (Adjusted \emph{R}\textsuperscript{2} = 0.175).  The additive regression model that explains exam points using practice and education is shown in Table \ref{tab:exam-points-practice-edu}. 

\begin{table}[htbp]
\centering
\caption{Explaining Exam Points Accounting for the Additive Relationship with Education Level}
\label{tab:exam-points-practice-edu}
\begin{tabular}{lS[table-format=1.2]cS[table-format=1.6,print-zero-integer=false]}
\toprule
\textbf{Explanatory Variables} & \textbf{Estimates} & $\mathbf{CI_{95\%}}$ & \textbf{p} \\ 
\midrule
Intercept & 0.05 & -0.49 -- 0.59 & .843 \\
\textbf{Doing} & \textbf{0.37} & \textbf{0.25 -- 0.49} & \textbf{$<$.001\threeS} \\
P7 or below & -0.27 & -0.86 -- 0.32 & .365 \\
S1 & -0.62 & -1.30 -- 0.06 & .074 \\
S2 & 0.12 & -0.53 -- 0.78 & .711 \\
S3 & -0.04 & -0.70 -- 0.62 & .907 \\
S4 & 0.13 & -0.45 -- 0.70 & .661 \\
S5 or above & 0.19 & -0.54 -- 0.91 & .612 \\
\midrule
\textbf{Observations} & \multicolumn{3}{c}{234} \\
\textbf{$R$\textsuperscript{2} / $R$\textsuperscript{2} adjusted} & \multicolumn{3}{c}{0.199 / 0.175} \\
\bottomrule
\end{tabular}
\end{table}

Exam score was better explained by variation between education levels. We conducted a likelihood ratio test to estimate, while accounting for education, which model explained more variance in exam scores. 
The likelihood ratio test revealed that Model 3 (interactive) described the variation in exam score better than Model 1 (additive), \emph{F}(6, 220) = 2.141, \emph{p} = 0.049, which suggests that the interaction between practice and education level significantly contributes to explaining differences in exam points across learners. 
We evaluated whether the relationship between doing and exam scores varies by education group. We found no significant interactions between doing and education levels ($p's > .05)$.
The lack of significant practice effect within each education group might be due to a lack of statistical power: the sample sizes within each group only included an average of $M = 33.43$ samples. Hence, our sample might not be large enough to reliably estimate and compare the relationship between practice and exam score within each education level; the interaction model better fits the data and better explains exam score outcomes. Still, we plot each group-level slope in Figure \ref{fig:interaction}, finding a descriptive trend toward a larger slope in lower education levels.

Figure \ref{fig:interaction} demonstrates that while increased practice is associated with improved exam score at all levels of education, the magnitude of this effect varies. 
For students at lower education levels, such as `P7 or below', the steep slopes indicate that practice has a more pronounced association with their exam scores. 
This suggests lower education learners may experience greater benefits from practice. Conversely, students at higher education levels (e.g., `Completed A Levels') start with a higher baseline of exam score and show a more modest increase with additional practice. 
This pattern suggests a ceiling effect, where students with more advanced prior knowledge benefit less from practice, likely because they are already closer to mastering the material. 
Thus, education level appears to moderate the effect of practice on performance, suggesting that prior knowledge plays a role in determining how much additional practice translates into knowledge benefits as assessed in the exam.

\textbf{\textit{RQ3: How does performance change with practice and listening to lecture? Is there a Doer Effect?}}
We estimated the relationship between completing more activities and the exam score considering prior listening to instruction. As shown in Table \ref{model-3}, doing remained significantly related to higher exam scores while accounting for listening, $CI_{95\%}$ = 0.42, $CI_{95\%}$ [0.26 – 0.50], \emph{p} <0.001. 
However, prior listening before practice alone did not relate significantly to exam score ($p's>~.05)$. \textit{Thus, we observe a Doer Effect.}

 There was a significant interaction between prior listening and practice when predicting exam outcomes. More prior listening was associated with an increase in the effect of practice on exam score, $\beta$ =  0.19, $CI_{95\%}$ [0.28 – 0.56], \emph{p}~<~.001. 
 We conducted a likelihood ratio test to compare the interactive and additive prior listening models. The likelihood ratio test indicated that the interactive prior listening model described the variation in exam score better than the additive model, \emph{F}(1, 220) = 5.51, \emph{p} = 0.020, suggesting the interaction between practice and the amount of prior listening contributes to explaining exam points. 

\begin{table}[htbp]
\centering
\caption{Effects of Doing and Prior Listening on Exam Score}
\label{tab:doingxlistening}
\begin{tabular}{lccc}
\toprule
\textbf{Explanatory Variables}       & \textbf{Estimates} &  $\mathbf{CI_{95\%}}$       & \textbf{p}                \\
\midrule
(Intercept)               & -0.02              & {[}-0.14 – 0.09{]} & .687                     \\
\textbf{Doing}                    & \textbf{0.42}               & \textbf{{[}0.28 – 0.56{]}}  & \textbf{\textless{}.001} \\
Prior Listening           & -0.02              & {[}-0.14 – 0.10{]} & .686                     \\
\textbf{Doing * Prior Listening }  & \textbf{0.18}               & \textbf{{[}0.03 – 0.33{]}}  & \textbf{.020}            \\
\midrule
\textbf{Observations}     & \multicolumn{3}{l}{234}                                             \\
\textbf{R\textsuperscript{2} / R\textsuperscript{2} adjusted} & \multicolumn{3}{l}{0.142 / 0.130}   \\         
\bottomrule
\end{tabular}
\end{table}

\section{Discussion}
Despite many prior demonstrations, the generalizability of the doer effect \cite{vancampenhoutDoerEffectScale2023, koedingerDoerEffectCausal2016} beyond WEIRD environments was still an open question. We provide initial evidence that the doer effect may generalize beyond WEIRD populations and to low-tech settings. We contribute insights from such a non-WEIRD environment to further the field's understanding of learning by doing across cultures and technology. Our findings may inform future curriculum and intervention designs of mobile learning.

Doing (active learning) corresponded with higher exam scores for these non-WEIRD learners in mobile learning. Despite the digital skills gap in Uganda \cite{world2019uganda}, learners used basic mobile technology to learn STEM concepts. This implies that there are learners who can leverage locally available technology for learning. Despite the technical limitations of basic phones, learners retained information throughout the multi-month program. Learning by doing in this environment motivates the delivery of remote STEM education to hard-to-reach populations through basic information technology. This is essential as lack of confidence, knowledge, and skills contribute to Uganda's low internet adoption rate as 90\% of households lack internet \cite{world2019uganda}.

Prior listening alone did not correlate with higher exam scores. However, learners who reported listening to radio lectures tended to practice more effectively.  Asher et al. \cite{asherPracticeFeedbackVs2024} noted that removing lectures undermined interest in course content for less-confident students, who may be discouraged when challenged to solve problems without upfront instruction. This contrast implies a need to understand the role of passive information sources in distance learning, particularly identifying for which learners they are most useful and when. 

We also identified variations of the Doer Effect between education levels and a potential ceiling effect limiting learners who might be willing to practice more. Practitioners facilitating educational programs in these environments may provide more active learning to benefit learners. Researchers may develop more holistic models of active learning by investigating the nature of active learning between education levels to understand their relation to the Doer Effect. Such efforts will require creative solutions to overcome the limitations of designing and evaluating learning activities with basic phones such as screen size, keypad interactions, and cellular service fees. 

\subsection{Future Work}
Learning analytics aims to create a loop where insights inform interventions and technology development that yield improved learning \cite{gasevicLearningAnalyticsShould2016}. We contribute to closing this loop by demonstrating that learners can generate learning metrics with basic ICT. 
Future work can focus on developing tracking and logging methods within radio and phone-based educational technology. Earlier designs of this study aimed to estimate the impact of radio listening by aligning broadcast times with the timing of learner responses. However, multiple broadcasts of lessons made it challenging to directly observe when, whether, and which broadcast learners listened to, which complicated isolating the effects of specific instructional design on student learning. How might future work overcome limitations to experimental control? How might basic mobile phones support interventions encouraging students to leverage the Doer Effect in their self-regulated learning? These capabilities may help reach learners more representative of the global population while adhering to the needs and affordances of all learners.  

\subsection{Limitations}
Our study offers one example of the Doer Effect in a non-WEIRD context, Uganda, demonstrating that the effect is not limited to WEIRD populations. We approach closing the gap in cross-cultural studies to understand where this effect generalizes. However, WEIRD is a spectrum of attributes, not a strict binary \cite{apicellaWEIRDReviewLast2020}, thus we cannot claim our findings generalize to every non-WEIRD environment. 

The radios and phones in our non-WEIRD context could not support all active learning activities studied with WEIRD populations, such as interactive simulations, drop and drag, and matching \cite{koedingerDoerEffectCausal2016}. However, learners performed cognitive tasks (e.g., information retrieval) similar to previous work. We execute a conceptual replication \cite{derksen2022kinds}, where previous theories and hypotheses are tested against slight variations in methodology, rather than a direct replication, where all procedures are strictly copied. 

\section{Conclusion}
The introduction of learning process data from the Majority World presents learners, researchers, and educators
many questions, such as how learners benefit from active learning, especially compared to passive learning in various technologies and demographics. Our study investigates rural Ugandan learners using radio and basic mobile phones as distance learning technology. The analyses presented in this paper contribute to a body of empirical knowledge and understanding of active and passive learning, particularly with basic mobile learning technology. Our findings suggest that active learning, beyond Western, Education, Industrial, and Democratic learning environments, is more positively associated with learning outcomes than passive learning, with some variation across education levels. This work offers a baseline for future studies that may investigate methods for capturing and aligning radio and phone learning events and may explore the impact of timing and combinations of active and passive learning for different types of learners.

 \begin{acks} 
 We thank the Jacobs Foundation for supporting
this study. We thank our partner, Yiya Solutions, Inc., who provided
access to the existing course data and supplementary material for our research. 
We also express our deepest gratitude to the students and
instructors of the studied course; this work would not have been
possible without their participation.
\end{acks}

\bibliographystyle{ACM-Reference-Format}
\bibliography{main}

\end{document}